\documentclass[12pt, a4paper]{article}
\reversemarginpar
\usepackage{ucs}
\RequirePackage{cmap}
\usepackage{authblk}
\usepackage[normalem]{ulem}
\usepackage{setspace}
\usepackage{lineno}
\usepackage{cite}
\usepackage{cleveref}
\usepackage[utf8x]{inputenc}  
\usepackage[T1, T2A]{fontenc}      
\usepackage[russian, english]{babel} 
\usepackage{xcolor}
\usepackage{amsmath}
\usepackage{amsfonts}
\usepackage{amssymb}
\usepackage[pdftex]{graphicx}
\usepackage{subfig}
\usepackage{tabularx}
\usepackage{upgreek}
\usepackage{textcomp}
\usepackage{tikz}
\usepackage{footmisc}

\usepackage[pagebackref=false]{hyperref}
\hypersetup{
    colorlinks,
    citecolor=blue,
    filecolor=blue,
    linkcolor=blue,
    urlcolor=blue
}

\usepackage{epstopdf}
\title{Light yield and radiation hardness studies of scintillator strips with a filler}

\author[1,a]{A.~Artikov}
\author[1]{V.~Baranov}
\author[1]{Yu.~Budagov}
\author[2]{M.~Bulavin}
\author[1,b]{D.~Chokheli\footnote{Corresponding author, email
 \href{mailto:chokheli@jinr.ru}{chokheli@jinr.ru} }}
\author[1]{Yu.I~Davydov}
\author[1]{V.~Glagolev}
\author[1]{Yu.~Kharzheev}
\author[1]{V.~Kolomoets}
\author[1]{A.~Simonenko}
\author[1,c]{Z.~Usubov}
\author[1]{I.~Vasiljev}

\affil[1]{\textit{Joint Institute for Nuclear Research, Dzhelepov Laboratory of Nuclear Problems, Dubna, Russia}}
\affil[2]{\textit{Joint Institute for Nuclear Research, Frank Laboratory of Neutron Physics, Dubna, Russia}}
\affil[a]{\textit{On leave of absence from Nuclear Physics Laboratory, Samarkand State University,Samarkand, Uzbekistan}}
\affil[b]{\textit{On leave of absence from High Energy Physics Institute, Iv. Javakhishvili Tbilisi State University, Tbilisi, Georgia}}
\affil[c]{\textit{On leave of absence from Institute of Physics, Baku, Azerbaijan}}

\begin{document}
	
\maketitle
\newpage
\begin{abstract}
\doublespacing
Detectors based on polystyrene scintillator strips with WLS fiber readout are widely used to register charged particles in many high-energy physics experiments. The fibers are placed into grooves or holes along the strip. The detection efficiency of these devices can be significantly increased by improving the optical contact between the scintillator and the fiber by adding an optical filler into the groove/hole. 

This work is devoted to the study of the light yield of a 5-m-long scintillator strip with a 1.2-mm-diameter Kuraray Y11(200)~MC WLS fiber inserted into the strip's co-extruded hole filled with synthetic silicon resin SKTN-MED(E). The light yield was studied using cosmic muons and a $^{60}Co$ radioactive source.

Radiation hardness study of viscous fillers and short strip samples were performed on the IBR-2 pulsed research reactor of fast neutrons at JINR.
\end{abstract}

\newpage
\tableofcontents
\noindent\rule{\hsize}{0.4pt}
\doublespacing
\section{Introduction}
Long scintillator strips are used in many high-energy physics experiments. The light collection for these detectors is usually performed by WLS fibers inserted in grooves or holes of the strips. For instance, in some experiments \cite{MINOS, OPERA, T2K}, to establish the best efficient coupling of the WLS fiber to the strip and to increase the light yield, the fiber was glued into the groove of the strip with an optical glue having a high transparency and a refractive index close to the refractive index of the strip base (usually polystyrene). 

To fill with glue the strip hole with a WLS fiber in it is a quite complicated task, especially for long strips. Therefore, the WLS fiber is usually inserted dry into the strip holes. For instance, in the Cosmic Ray Veto (CRV) system for the upcoming Mu2E experiment (Fermilab, \cite{mu2e}), WLS fibers will mostly be inserted dry into the strip holes. In this case, light propagation from the scintillator to the WLS fiber is through the air layer. Large difference in refractive indices at the "scintillator-to-air" and at "air-to-outer WLS fiber cladding" interfaces results in high losses of light because of scattering. Injection of some optically transparent liquids (fillers) into the strip holes led to the light yield growth up to 50\% in comparison to the "dry" strips \cite{pepan2017}. 

A low-molecular synthetic resin SKTN-MED(E) \cite{SKTN} revealed good gain in light yield, so we chose it for further studies. However, the resin base has a high viscosity ($10-20 \: Pa*s$) and, therefore, its injection into the strip hole 2.6~mm in diameter with a 1.2-mm-diameter WLS fiber already installed in it was a complicated task. We developed a special technique to solve the problem \cite{pepan2017,jinst2016}. Light yield study of a 2-m-long polystyrene strip (with dopants 2\% РТР and 0.03\% РОРОР \cite{ISMA}) with a 1.2~mm-diameter Kuraray Y1(200)MC WLS fiber \cite{Y11} filled with the SKTN-MED(E) silicon resin base revealed a factor of 1.6-1.9 increase in the light yield in comparison to the same strip but with no filler in it. This strip had a triangular cross-section with a base of 33~mm and a height of 17~mm and a 2.6-mm-diameter co-extruded hole.

It was of great practical interest to verify applicability of the developed filling technique to long strips and study their light collection. For this purpose, a 5-m-long strip was assembled \cite{Minsk2016}, and its light yield was studied using cosmic muons and a $^{60}Co$ radioactive source.

\begin{figure}[!htbp]
	\begin{center}
		\includegraphics[height=0.15\textheight]{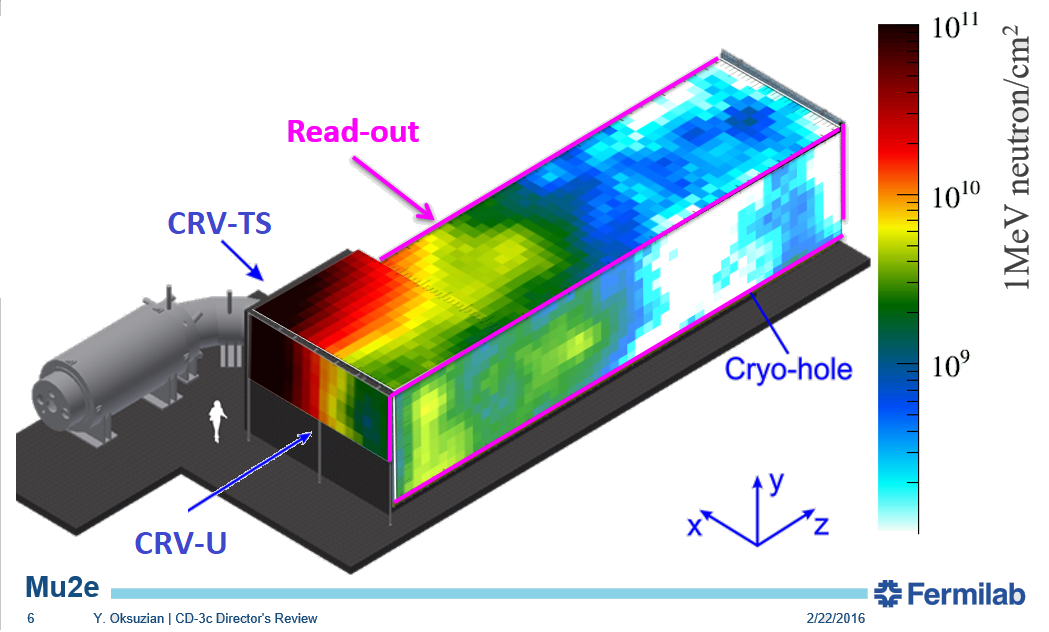}
		\caption{\small Expected distribution of neutron fluxes \cite{mu2e_oks}}
		\label{fig: mu2eneutrons}
	\end{center}
\end{figure}

Long-term stability of parameters of detectors based on scintillator strips is important for experiments at modern accelerators. Deterioration of these parameters may be due to influence of radiation and natural factors (temperature, humidity). In particular, part of the CRV system (for mu2e experiment at FNAL, \cite{mu2e}), which is a multilayer array of strips with WLS fibers in them, will experience significant radioactive exposure to neutron fluxes during the data taking (see Figure~\ref{fig: mu2eneutrons}, \cite{mu2e_oks}). Six CRV modules close to the transport solenoid (CRV-TS) and three CRV modules in front of the pipeline wall (CRV-U) will be irradiated by the highest expected neutron flux among the CRV modules, up to $10^{11}$ neutrons per cm$^2$ \cite{mu2e_oks}.

To investigate the effect of this type of radiation, we studied radiation hardness of SKTN-MED (grade E and D) fillers, epoxy resin BC-600 \cite{Bicron}), and short strip samples with these fillers on the IBR-2 pulsed research reactor of fast neutrons (FLNP, JINR, \cite{IBR2}).

\label{introduction}

\section{Selection of an optical filler and its injection into a strip}
The filler, as an intermediate medium transferring light from a scintillator to a WLS fiber, should have a number of properties: high optical transparency, refractive index close to the refractive index of polystyrene, good adhesion to the scintillator, and long-term stability of parameters.

Several different substances (water, aqueous solution of glycerin, UV-adhesive SPECTRUM K-59EN \cite{K59H}, SKTN-MED (E)) were tested as fillers for short (50 ~ cm) strips \cite{pepan2017}. The first three {low-viscosity substances ($<20 \: mPa*s$)} were injected into the strip by a syringe, but a special technique was required for injection the high-viscosity SKTN-MED(E) resin ($ 10-20 \: Pa * s $). This technique was tested on a short (50 cm) strip \cite{pepan2017}. The study of the light yield of the 50-cm-long strips with these optical fillers showed a 50\% increase in the light yield against the "dry" strip. The best result was obtained with the resin.

Full-scale studies with this filler were carried out \cite{jinst2016}. SKTN-MED(E) has high transparency in the visible region of the spectrum (about 95\%). A refractive index is 1.40\footnote{A wrong value of the refractive index was given in our previous publications \cite{pepan2017} cited from \cite{ISMA} }  (see Table \ref{table: refindex}), which is very close to the WLS outer layer refractive index of 1.42.
This is very important since the light undergoes scattering at the interface of two media with refractive indices $n_{1}$ and $n_{2}$, and the reflection coefficient for the light rays normally incident on the interface is described by the Fresnel formula
\begin{equation} \label{eq:Fresnel}
R = \left[ \frac{n_{1}-n_{2}}{n_{1}+n_{2}}\right] ^2
\end{equation}

\begin{table}[!htbp]
	\footnotesize
	\begin{center}
		\begin{tabular}{|c|c|}
			\hline
			Filler & Refractive Index\\
			& measured at 25\textdegree{C} \\
			\hline
            Water, distilled & 1.330 \\
			\hline
            Ethanol & 1.3617 \\
			\hline
            PMX-200/1000 \cite{pmx200} & 1.4025\\
			\hline
            SKTN-D & 1.4035\\
			\hline
            SKTN-B & 1.4035\\
			\hline
            BC600 & 1.568\\
			\hline
		\end{tabular}
	\end{center}
	\caption{Refractive indices for different fillers measured at 25\textdegree{C} with the refractometer IRF-454 B2M \cite{irf454} at the Department of Chemistry, New Technologies and Materials, Dubna State University, Russia} 
	\label{table: refindex}
\end{table}

\begin{figure}[!t]
  \centering
   \subfloat[Layout of the filling setup]{\includegraphics[height=0.1\textheight, keepaspectratio] {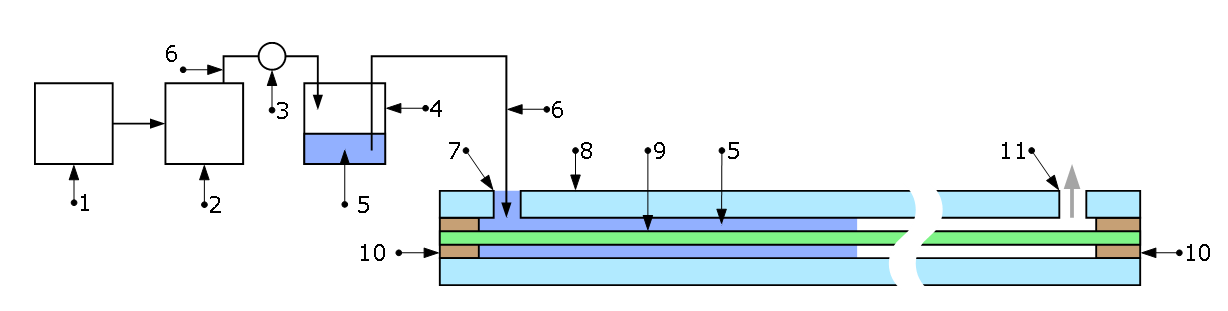}}
   \hfill
   \subfloat[Drechsel bottle]{\includegraphics[height=0.1\textheight, keepaspectratio] {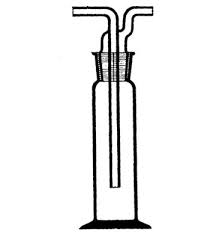}}
   \captionsetup{singlelinecheck=off}
   \caption{\small Layout for the setup to pump the high viscosity filler into the co-extruded hole of the scintillator bar (no scale):
   	\\ a) (1) dry type compressor; (2) SL101N Digital Liquid Dispenser; (3) manometer; (4)~special vessel with filler; (5) filler; (6) polyvinylchloride tube; (7) inlet for filling; (8) strip; (9) WLS fiber; (10) sealing; (11) exhaust outlet for extracting air.
   	\\ b) Drechsel bottle with inlet and outlet and cap} 
 	\label{fig: setup4filling}
 \end{figure}

The setup used to insert the SKTN-MED(E) filler into the strip was upgraded in comparison with that used in \cite{pepan2017, jinst2016}. The layout of the pumping setup is shown in Figure \ref{fig: setup4filling}. The dry-type compressor (1) produces initial pressure, and the SL101N Digital Liquid Dispenser (2) provides a constant pressure at the output (about 0.2–0.4 bar above atmospheric pressure), which is monitored by the manometer (3). This constant pressure is provided in a special vessel based on the Drechsel bottle (4) with an optical filler in it (5) and forces the filler to flow through the tube (6) and the inlet (7) into the hole of the strip (8) with the WLS fiber (9) in it.

Both ends of the WLS fibers (10) are glued by 5-min transparent epoxy glue (Hardman RED 04001 \cite{RED04001}) to the strip edges before the filling procedure (10). Air is removed from the strip hole through the exhaust outlet (11). Once filling is done, the inlet and outlet holes of the strip are sealed with 5-min transparent epoxy glue.

We filled the 5-m-long strip with the WLS fiber by SKNT-MED(E). The cross-section of this strip was the same as in \cite{jinst2016}, with a base of 33 mm and a height of 17 mm and a hole 2.6 mm in diameter at the center. The filling time was about six hours, or 2.5 times longer than the filling time for the 2-m-long strip \cite{jinst2016} under same conditions. Later on \cite{UVAstrips2016}, we improved the filling technology, so that the filling procedure was carried out simultaneously for eight holes of 4.5-m-long strips. The average injection time was about 2.5 h in the case of using SKTN-MED(D). The excess pressure was maintained at 8 psi. Note that viscosity of this filler is two times lower than that of SKTN-MED(E). 
\label{filling}

\section{Light yield of a stintillator strip}
We studied the light yield of a 5-m-long strip as a function of the distance from the PMT using cosmic muons and radioactive sources. In addition, MC simulation of the light collection of a 5-m-long strip was done.
\label{lightcollection}

\subsection{Light yield modeling}\label{model}
MC simulation of the light collection was conducted using Geant\,4, version 10.3.1. The geometry of the 5-m-long scintillator strip with a WLS fiber in it and a reflective coating described by Geant\,4 corresponded to the actual samples used in the measurements. The cosmic muon energy was modeled according to \cite{volk, volk1}.

Muons generated by this model were normally incident on the strip at one point at a certain distance from the strip end where the photodetector should be installed.

\begin{figure}[!htbp]
	\centering
	\subfloat[Simulated light yield from the 5-m-long strip with a triangular cross-section under different conditions.]{\includegraphics[height=0.15\textheight]{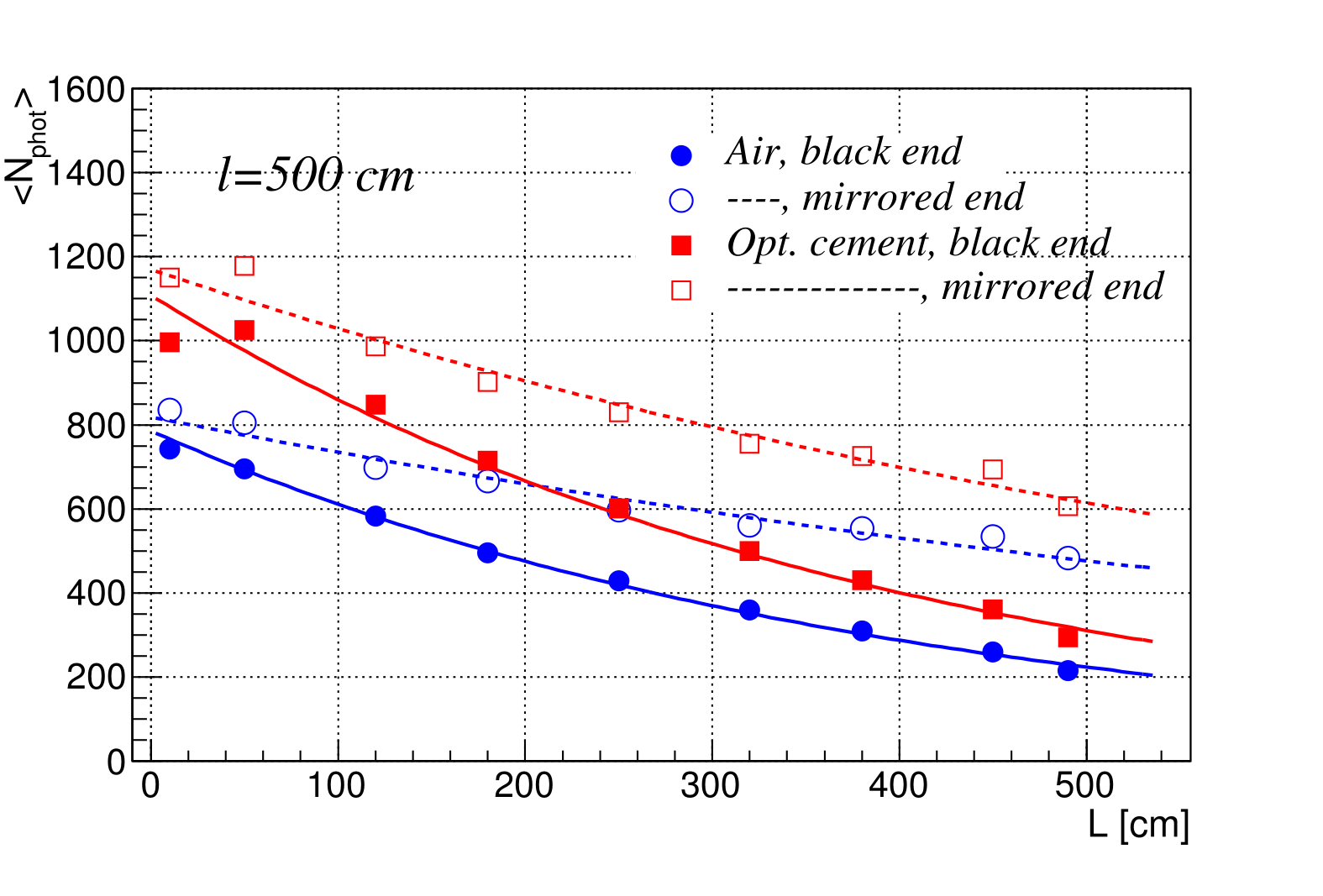} \label{AttenAbs}}
	\hfill
	\subfloat[Relative increase in the light yield.]{\includegraphics[height=0.15\textheight]{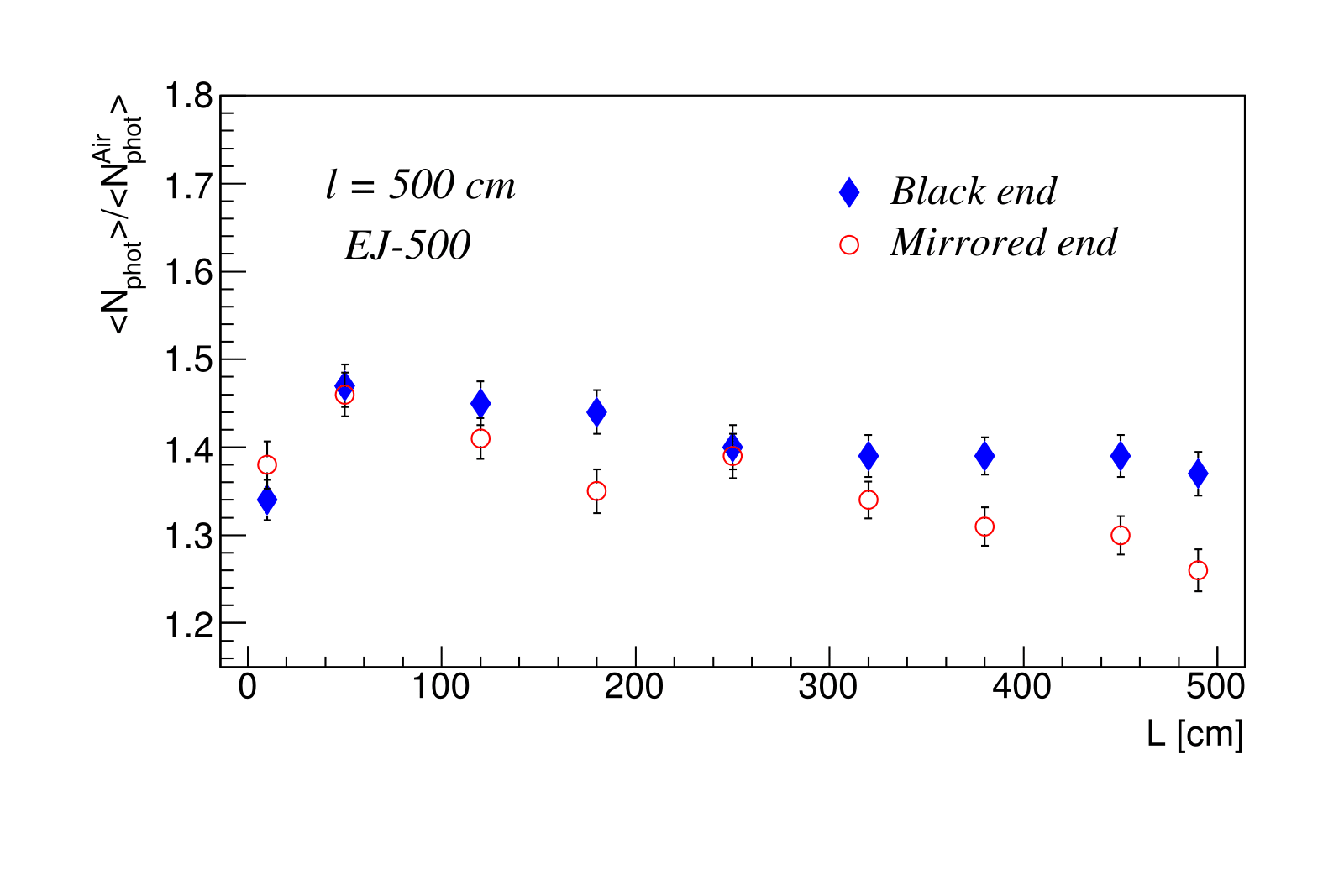}\label{AttenRel}}
	\caption{\small Simulation results for the light yield from the 5-m-long strip with a triangular cross-section.}
	\label{fig: Atten}
\end{figure}

Light emission initiated by the passage of the muons through the scintillator, light absorption and subsequent re-emission by the WLS fiber, and light reflection and refraction on the fiber and strip coating were simulated according to the chosen optical model described in Geant\,4. While modeling, we limited ourselves to counting photons at the end of the fiber and did not consider the processes associated with the photodetector. Dependence of the average number of photons arriving at the WLS fiber end on the distance to the point of incidence of cosmic muons is shown in Figure~\ref{AttenAbs}. The points in this figure correspond to different ways of processing the far end of the fiber and to the cases without a filler and with an optical filler with properties similar to those of SKTN-MED(E):

\begin{itemize}
	\item[] a strip model with no filler and the blackened far end (blue filled circles);
	\item[] a strip model with no filler and the mirrored far end (blue open circles);
	\item[] a strip model with a filler and the blackened far end (red open squares);
	\item[] a strip model with a filler and the mirrored far end (red filled squares).
\end{itemize}

The curves in Figure \ref{AttenAbs} correspond to the approximation of the points by an exponential function.

The effect of the optical filler on the average number of photons for each point is shown in Figure \ref{AttenRel}. The ratio between the average numbers of photons for the filled and unfilled scintillator bars shows a significant advantage of the proposed method for increasing the light collection.

\begin{figure}[!htbp]
	\centering
	\subfloat[Far end of the WLS fiber blackened]{\includegraphics[height=0.15\textheight]{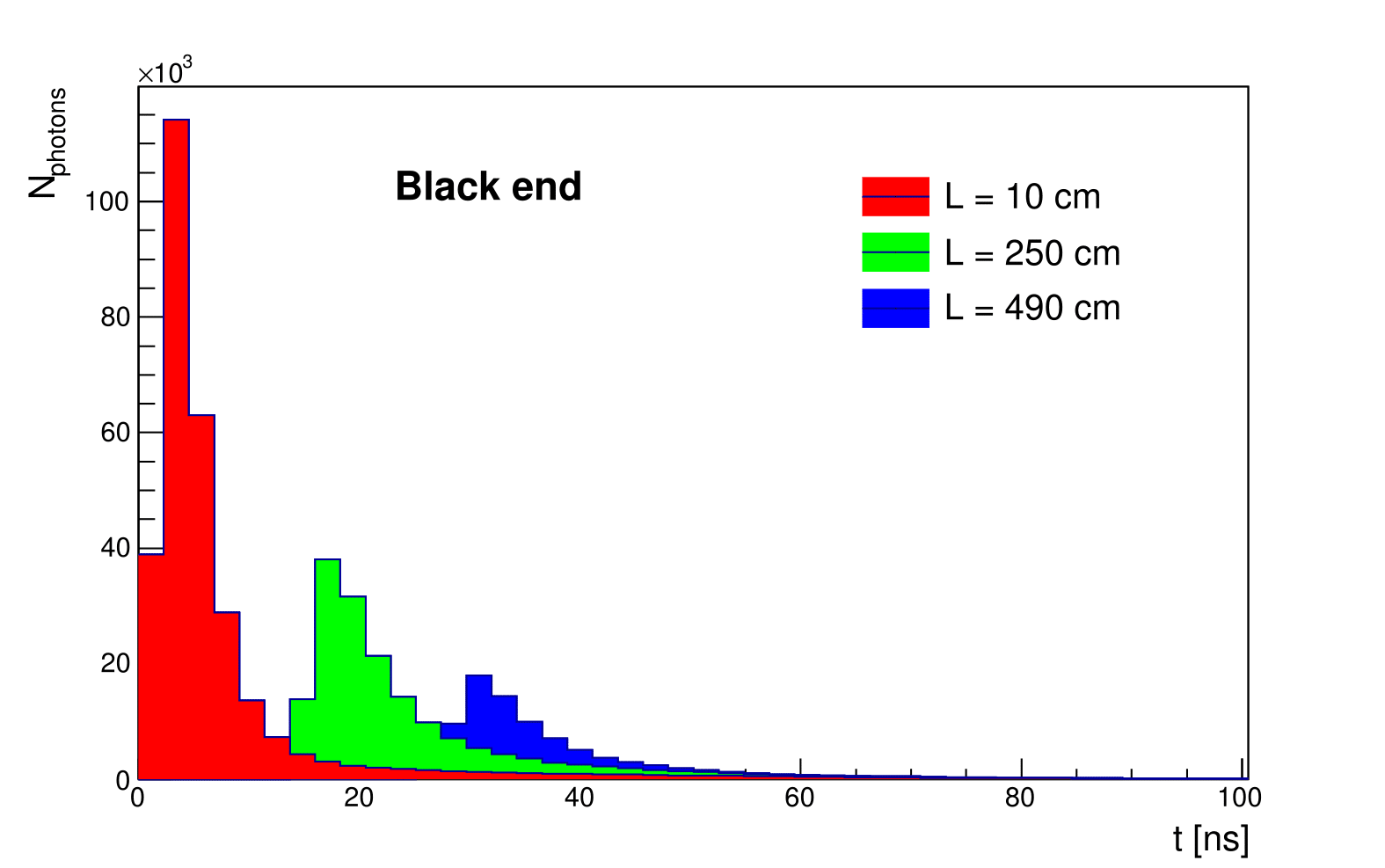} \label{TimeDark}}
	\hfill
	\subfloat[Far end of the WLS fiber mirrored; reflected signals appear]{\includegraphics[height=0.15\textheight]{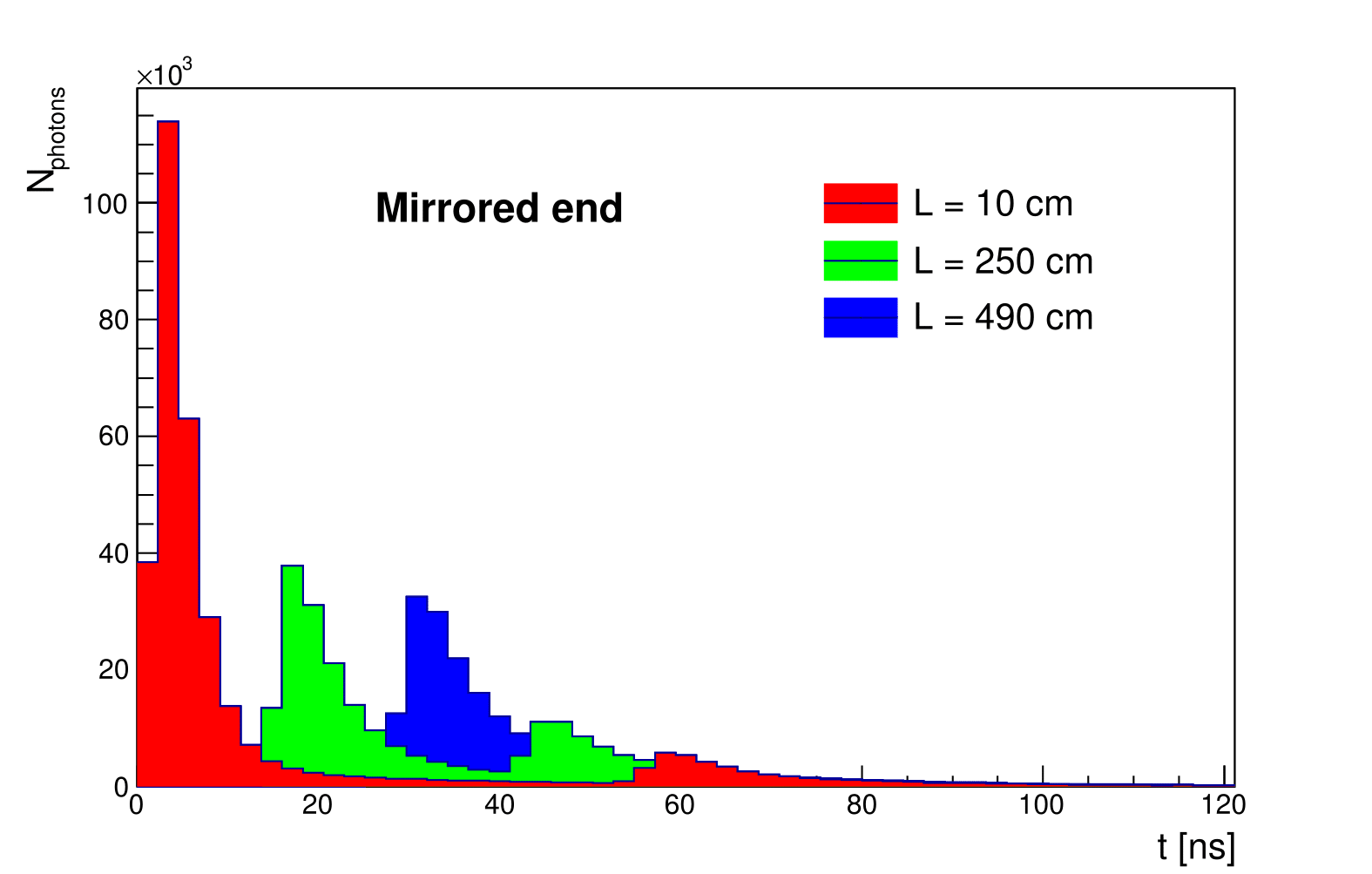}\label{TimeMirror}}
	\caption{\small Simulation results for the light signal coming to the photodetector from different distances}
	\label{fig: Time_dis}
\end{figure}

The dispersion of the time for photons to reach the photodetector surface and the signal intensity significantly depend on the reflectivity of the far end of the fiber. Figure \ref{fig: Time_dis} shows dependence of the number of photons on the time of their reaching the photodetector for different distances from the point of incidence of cosmic muons and for the blackened and mirrored ends of the fiber.

With the far end mirrored, the delayed signal of the photons reflected from the far end of the WLS fiber was clearly separated (Figure \ref{TimeMirror}). This fact indicates that the mirroring leads to light yield increase but deteriorates the signal time resolution and should not be recommended for long scintillator strips if the arrival time of the reflected signal exceeds the designed triggering gate of the data acquisition system.

\begin{figure}[!htbp]
	\begin{center}
		\includegraphics[height=0.15\textheight]{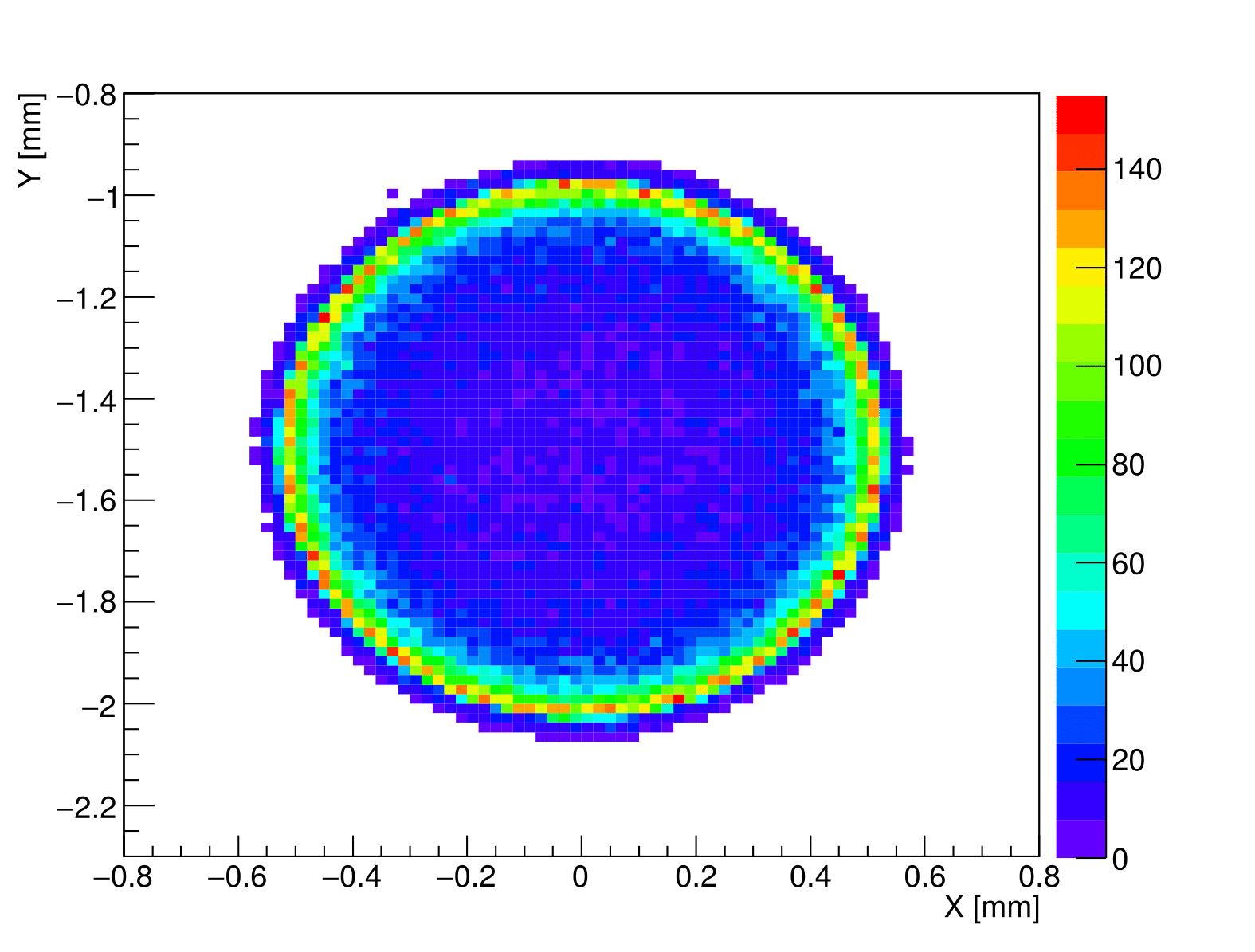}
		\caption{\small Light distribution at the WLS fiber end}
		\label{fig: Fiber_CS}
	\end{center}
\end{figure}

Knowledge of the light intensity distribution by the WLS fiber cross section plays an important role in selection of the optimal photodetector. The simulation results (see Figure~\ref{fig: Fiber_CS}) show that the light intensity in the optical fiber increases radially from the center outward and reaches the maximum in the region close to the first fiber cladding.

\subsection{Light yield study with cosmic rays}\label{cosmic}
\begin{figure}[!htbp]
  \centering
   \subfloat[]{\includegraphics[height=0.14\textheight, keepaspectratio]
			{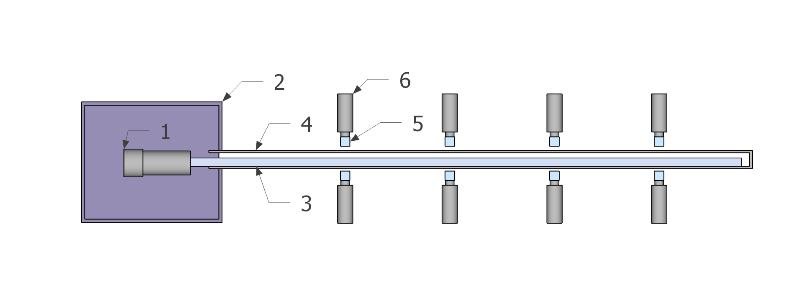}}
  \hfill
  \subfloat[]{\includegraphics[height=0.1\textheight, keepaspectratio]
			{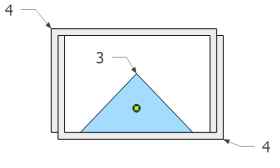}}
  \caption{\small Layout for the experimental setup (a), cross section of light-proof Al U-channels with a strip inside (b) (1) EMI9814B PMT, (2) black box, (3) strip, (4) lightproof Al U-channel, (5) four pairs of trigger scintillation counters $20\times25\times30 \: mm^3$, (6) FEU 85 PMT.}
	\label{fig: cosmicsetup}
\end{figure}

Light yield study of the 5-m-long strip using cosmic muons was carried out by the same method as used for the 2-m-long strips \cite{jinst2016} (see Figure~\ref{fig: cosmicsetup}). The measurements were done at different distances from the PMT and under various conditions (a strip filled with SKTN-MED(E) and with no filler; the strip and the WLS fiber ends far from the PMT blackened or covered with aluminized mylar). Light was collected by the EMI9814B PMT (1) \cite{EMI9814} placed inside of the lightproof box (2). The readout strip (3) was placed inside of two lightproof Al U-channels (4) covered by black paper inside (not shown). The cosmic-ray telescope was comprised of four pairs of plastic scintillation counters (5) with dimensions of $20\times25\times30 \: mm^3$ coupled to the FEU85 PMT (6) \cite{PMT85}.

\begin{figure}[!htbp]
	\begin{center}
		\includegraphics[height=0.14\textheight, keepaspectratio]
		{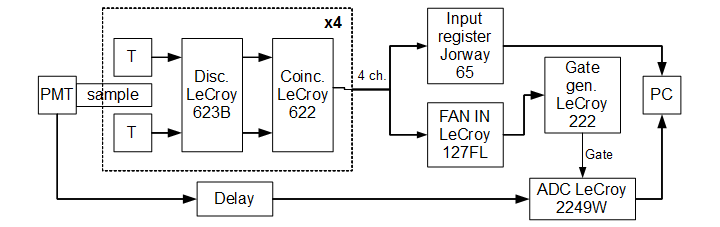}
		\caption{\small  Block diagram of the electronics used for measurements with cosmic muons}
		\label{fig: cosmicblock}
	\end{center}
\end{figure}

The DAQ system for the experiment setup was made  using the NIM and CAMAC modules. The electronics block diagram of the experiment is shown in Figure \ref{fig: cosmicblock}. The analog signal from the PMT is measured by the LeCroy ADC~2249W charge-to-digital converter. Signals from the cosmic telescope are discriminated by the LeCroy~623B; the LeCroy~622 coincidence module creates an output signal to mark the position by the Jorway~65 input register and runs the LeCroy~222C gate generator to produce the strobe signal with a width of 100~ns.

\begin{figure}[!htbp]
	\begin{center}
		\includegraphics[height=0.2\textheight, keepaspectratio]
		{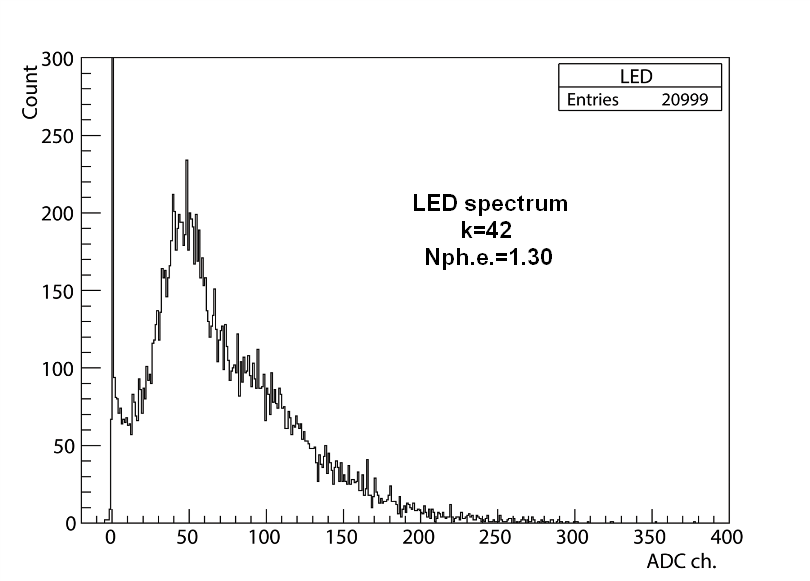}
		\caption{\small Typical calibration spectrum obtained with the LED.}
		\label{fig: ledspectra}
	\end{center}
\end{figure}

The spectrometric channel was calibrated using a PMT single electron peak \cite{abscal}. The calibration was performed with a “NICHIA” NSPB310A light emission diode (LED) \cite{LED} using flashes of low-intensity light sent to the photocathode of the EMI9814B PMT. One of the typical calibration spectra obtained with the LED is shown in Figure~\ref{fig: ledspectra}, where one photoelectron peak is clearly observed. It is possible to distinguish the second peak as well.

\begin{figure}[!htbp]
	\centering
	\subfloat[]{\includegraphics[height=0.2\textheight, keepaspectratio]
		{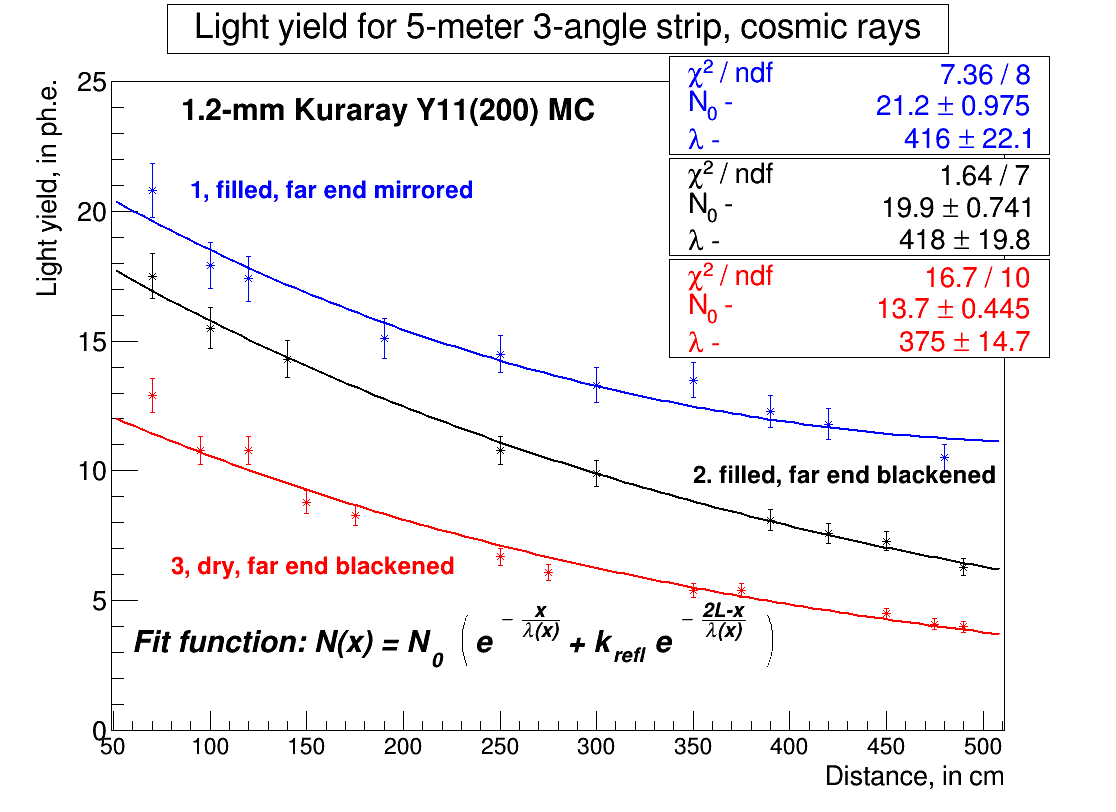}\label{cosmiclightall}}
	\hfill
	\subfloat[]{\includegraphics[height=0.2\textheight, keepaspectratio]
		{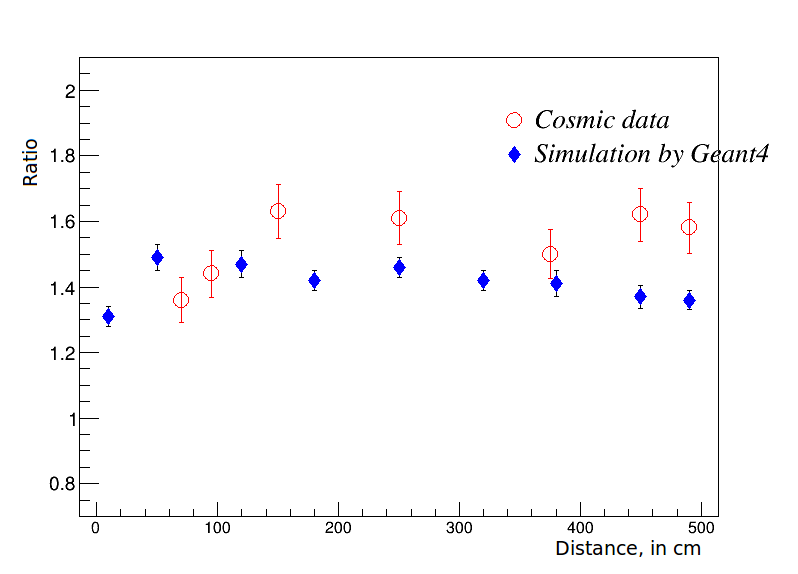}\label{cosmiclightratio}}
	\caption{\small (a) Light yield (in photoelectrons) of the 5-m-long strip hit by comic muons under different conditions of the study: strip filled with SKTN-MED, far-from-PMT end mirrored (curve 1) or blackened (curve 2); strip without a filler, far-from-PMT end blackened (curve 3). \\(b) Correlation of the experimental data with the simulation.}
	\label{fig: cosmiclight}
\end{figure}

Figure \ref{fig: cosmiclight} presents the results of measuring the light yield of the strip with the 1.2-mm-diameter Kuraray Y11(200) MC fibers and SKTN-MED(E) filler; the far-from-PMT end of the strip and the WLS fiber were covered by mylar (curve 1) or blackened (curves 2 and 3). The experimental data  were fitted with biexponential functions \cite{Minsk2016, UVaOptStudy}
 \begin{equation} \label{eq:TALcos}
 N(x) = N_{0}\left(e^{ - \frac{x}{\lambda(x)}} + f_{Ref} e^{ - \frac{2L-x}{\lambda(x)}}  \right)
 \end{equation}
 where $x$ is the distance along the strip; $N_{0}$ is the initial amount of the light yield; $\lambda(x)$ is the technical attenuation length; $L$ is the length of the strip, and  $k_{refl}$ is the reflectance of the surface of the material used as a mirror (about 75\% for the aluminized mylar reflector and less than 5\% for the blackened end). 

For direct comparison of the light yield from the strip with and without a filler the light yield of the “dry” strip was studied first. The light yield increases in the strip with a filler and a blackened end in comparison with the "dry" strip (approximately by 50\%) (curves 2 and 3). A much higher increase in the light yield (up to a factor of two) is observed in the strip with the filled hole when the far-from-PMT ends of the strip and the WLS fiber are covered by aluminized mylar.

Increase in the light yield of the strip filled with SKTN-MED(E) against the light yield of the same strip but without a filler as a function of the distance to the PMT is presented in Figure \ref{cosmiclightratio}; the far-from-PMT end was blackened. The simulation generally agrees with the experimental data.   

Mirroring the far-from-PMT end of the fiber gives a gain in the light yield, but the broadening of the light signal in the arrival time to the photodetector appears (see chapter \ref{model}). Therefore, when the temporal resolution is crucial, the far end of the strip (fiber) should be blackened. The speed of light in the WLS fiber is around 16~cm/ns \cite{Miniskirt, Mineev}. This will result in a delay of the light signal reflected from the far end of the strip by about 40~ns for the passing muons in the middle of the 6.5-m-long strip (see also Figure \ref{TimeMirror}); such a strip corresponds to the longest detectors to be used for the CRV system in the mu2e experiment.

\subsection{Light yield study with a $^{60}Co$ radioactive source}\label{radsource}
A radioactive $^{60}Co$ source was used to study the light yield of the 5-m-long strip as a function of the distance to the PMT by measuring the anode current of the EMI9814B PMT using the Keitley 6487 picoammeter \cite{K6487}.

The source was placed on a special support which was moved along the strip.

\begin{figure}[!htbp]
	\begin{center}
		\includegraphics[height=0.20\textheight]{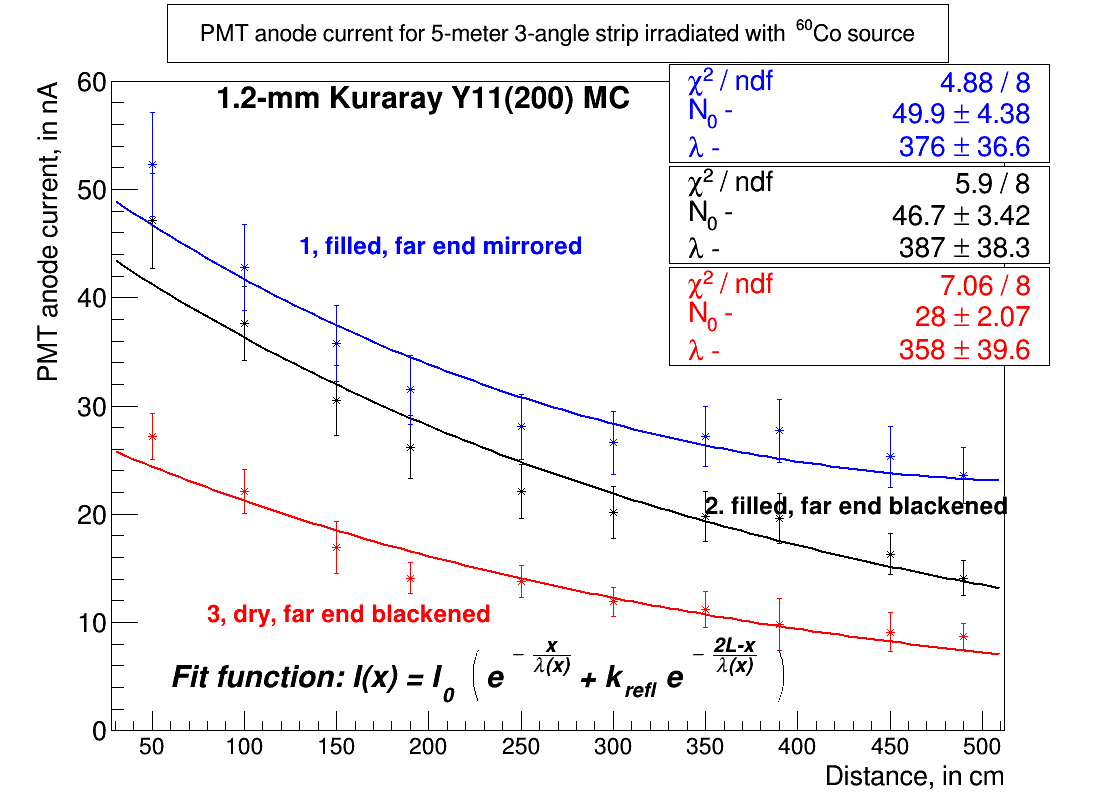}
		\caption{\small Light yield (anode current in nA measured by the picoammeter) as a function of the distance from the PMT for the 5-m-long strip irradiated by the radioactive $^{60}Co$ source. Two upper curves correspond to the strip with the SKTN-MED(E) filler and the lower curve corresponds to the same strip without a filler. The far-from-PMT ends of the WLS fiber and the strip were covered by Al mylar (curve 1) or blackened (curves 2 and 3).}
		\label{fig: lightradio}
	\end{center}
\end{figure}

The results of the study are shown in Figure~\ref{fig: lightradio}. The experimental data  were fitted with biexponential functions similar to Equation~\ref{eq:TALcos}, but for the PMT anode current instead of the number of photoelectrons
\begin{equation} \label{eq:TALrad}
I(x) = I_{0}\left(e^{ - \frac{x}{\lambda(x)}} + k_{refl} e^{ - \frac{2L-x}{\lambda(x)}}  \right)
\end{equation}

The upper two curves correspond to the strips filled with SKTN-MED(E) and the lower ones correspond to the strips with no filler; the far-from-PMT ends of the strip and the WLS fiber were mirrored (curve~1) or blackened (curves 2 and 3).  The light yield of the strip with the filler is at least 1.5 times higher than the light yield of the strip without a filler.

Using mylar with aluminized coating as a mirror at the far-from-PMT end of the strip with the WLS fiber gives an increase in the light yield by almost a factor of two in comparison to the case where the far end of the strip was blackened; this comparison study was performed only for the "dry" strip. This result emphasizes importance of mirroring this strip end when temporal resolution is not crucial.

\section{Radiation hardness study}\label{radhard}
Long-term stability of properties of detectors based on scintillator strips becomes particularly important 
under their continuous operation in modern experiments; these properties can deteriorate under the influence of both natural factors (temperature, humidity, time) and radiation. Deterioration of properties or "aging" of scintillator detectors can occur due to degradation of the scintillator base (polystyrene) and the dopants and due to degradation of the WLS fiber and the filler. Some aspects of radiation hardness and natural aging studies are published in \cite{zorn93, britvich93, radneutron, vara00, radwls91, scintUPS923a, radcms, scintCDFaging, Grinev, Senchishin}.

\subsection{Radiation hardness study on the IBR-2 pulsed research reactor of fast neutrons (FLNP, JINR)}\label{irb}
The IBR-2 pulsed research reactor of fast neutrons \cite{IBR2nimb} (FLNP, JINR) was used to research radiation damage of strips and synthetic silicon resins SKTN-MED(D) and SKTN-MED(E).

The SKTN-MED(D) and SKTN-MED(E) resins and short polystyrene strip samples with the WLS fiber filled or not filled with those resins were irradiated by a neutron beam of various density.

Study of the 2-m-long strips with the SKTN-MED(E) and BC-600 fillers showed very close results in light yield \cite{pepan2017, jinst2016}; therefore, it was of particular interest to compare radiation hardnesses of these fillers. We irradiated strips with the BC-600 filler based on epoxy resin as well. This filler had more than 98\% transparency in the visible region of the spectrum and the refractive index of 1.56.

The 2-m-long strip was cut into 12 samples 15 cm long to carry out radiation tests; inlet and outlet holes were drilled in each sample for pouring the filler in and letting the air out during the filling. The 1.2-mm-diameter Kuraray Y11(200) MC WLS fiber was inserted into the strips. The ends of the fibers were fixed with 5-min transparent epoxy glue (Hardman RED 04001 \cite{RED04001}) at the ends of the strips and polished. The strip ends were covered with black plastics with a hole to bring the fiber out.

These 12 short strips were divided into three groups. In each group one strip was "dry" and the other three were filled with SKTN-MED(E), SKTN-MED(D) or BC-600. Each of the fillers (base/resin) was poured into plastic containers (a total of nine containers). Then strips with different fillers and containers with different resin were placed at different distances from the reactor core. Thus, three different radiant exposures (fluences) of $16\times10^{14}$, $3.8\times10^{14}$, and $1.2\times10^{14}$ neutrons/cm$^2$ were provided (see Table \ref{table: radexpo}, which also presents the corresponding neutron flux densities and absorbed rate doses for $\gamma$ particles). The indicated fluxes refer only to fast neutrons ($E>1 \: MeV$), and the fluxes of slower (thermal and resonance) neutrons are not included in this table. Therefore, the real radiation doses received by the samples were somewhat higher.

\begin{table}[!htbp]
	\footnotesize
	\begin{center}
		\begin{tabular}{|c|c|c|c|}
			\hline
			Location & Flux density & Fluence  & Background of $\gamma$\\
			& $n/cm^{2}$c & $n/cm^{2}$ & $Mrad$ \\
			\hline
			1 & $1.8\times10^9$ &	$16\times10^{14}$	& 5.4 ... 1.4 \\
			\hline
			2 & $4.4\times10^8$	& $3.8\times10^{14}$	& 0.47 \\
			\hline
			3 & $1.35\times10^8$	& $1.2\times10^{14} $ &	0.37 \\
			\hline
		\end{tabular}
	\end{center}
	\caption{Neutron flux density, fluence, and $\gamma-$particles dose rate for the irradiated samples ($E>1 \: MeV$)}
	\label{table: radexpo}
\end{table}

\subsection{Optical properties of glues and their bases after irradiation on the IBR-2}\label{irbglue}
\begin{figure}[!htbp]
	\centering
	\subfloat[SKTN-MED(D)]{\includegraphics[height=0.2\textheight, keepaspectratio]
		{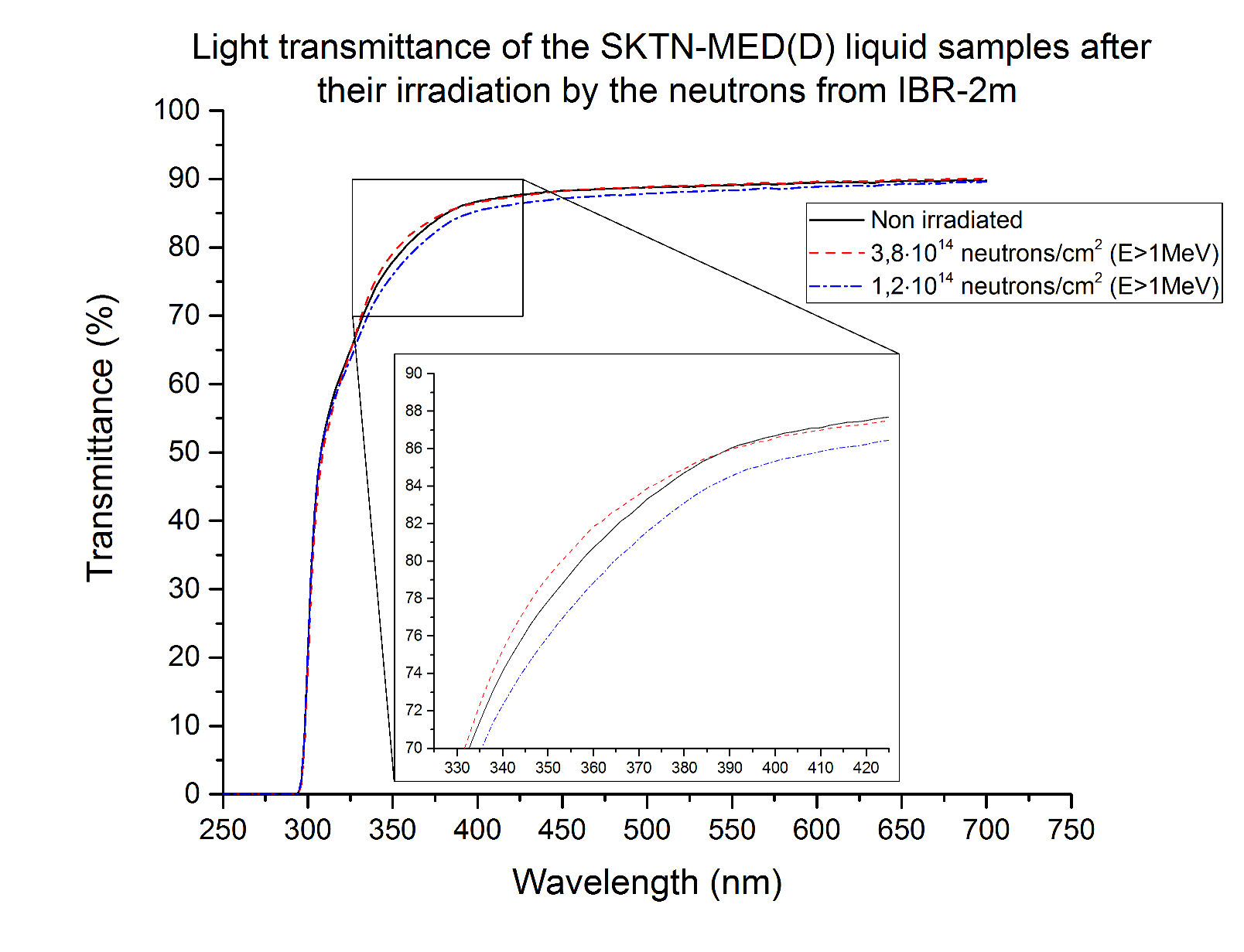} \label{fig: transpSKTNd} }
	\hfill
	\subfloat[SKTN-MED(E)]{\includegraphics[height=0.2\textheight, keepaspectratio]
		{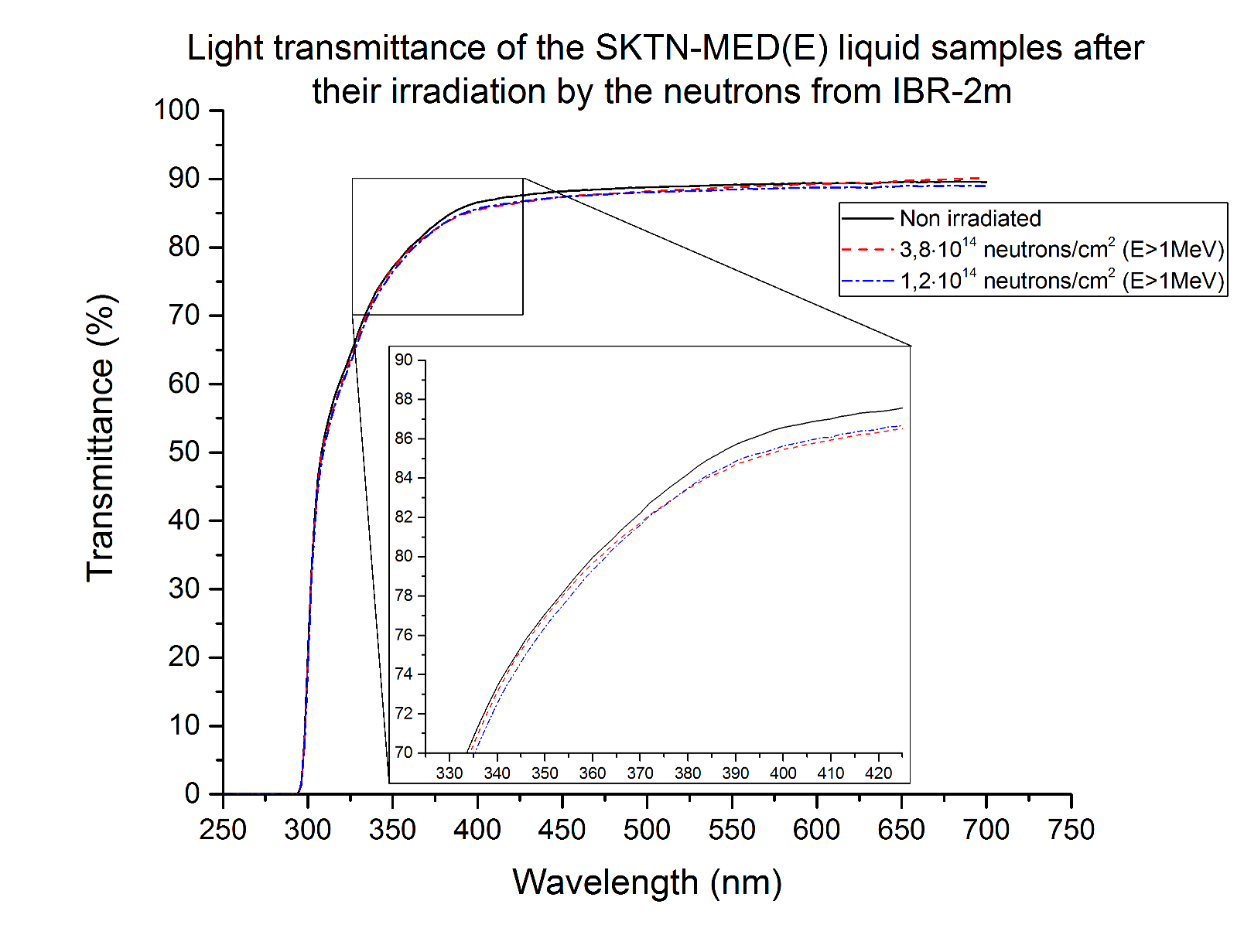} \label{fig: transpSKTNe} }
	\caption{\small 
		Transmittance of SKTN-MED(D) and SKTN-MED(E) as a function of the light wavelength before and after the irradiation by the neutron beam with the fluences of $3.8\times10^{14}$ and $1.2\times10^{14}$ neutrons per cm$^{2}$. The Plexiglas window suppressed strong ultraviolet light (<~300~nm).}
	\label{fig: transpSKTN}
\end{figure}

The transmittance of the glues and their bases in the light region of 200 to 800 nm before and after irradiation was measured by a spectrophotometer \cite{shimadzu}. The resin bases were poured into small cuvettes for measurements. The cuvettes were made on a 3D printer \cite{3Dprinter} providing the filler layer thickness of 4~mm. The windows of the cuvettes were made from a regular 1-mm-thick Plexiglas sheet, which led to the suppression of the ultraviolet region of the spectrum shorter than 300~nm. The results of these measurements for SKTN-MED(E) and SKTN-MED(D) are given in Figures \ref{fig: transpSKTNd} and \ref{fig: transpSKTNe}, respectively.

One can see that transmittance of both fillers is close to 90\% at the light wave length more than 400~nm after the irradiation by a neutron beam with the fluences of $3.8\times10^{14}$ and $1.2\times10^{14}$ per cm$^{2}$. This result is almost the same as before the irradiation. Note that the irradiation with the fluence of $16\times10^{14}$ neutrons per cm$^{2}$ led to polimerization of the filler (resin), which made it impossible to pour this filler into the cuvette and measure its  transmittance by the spectrophotometer.

\begin{figure}[!htbp]
	\begin{center}
		\includegraphics[height=0.2\textheight, keepaspectratio]
		{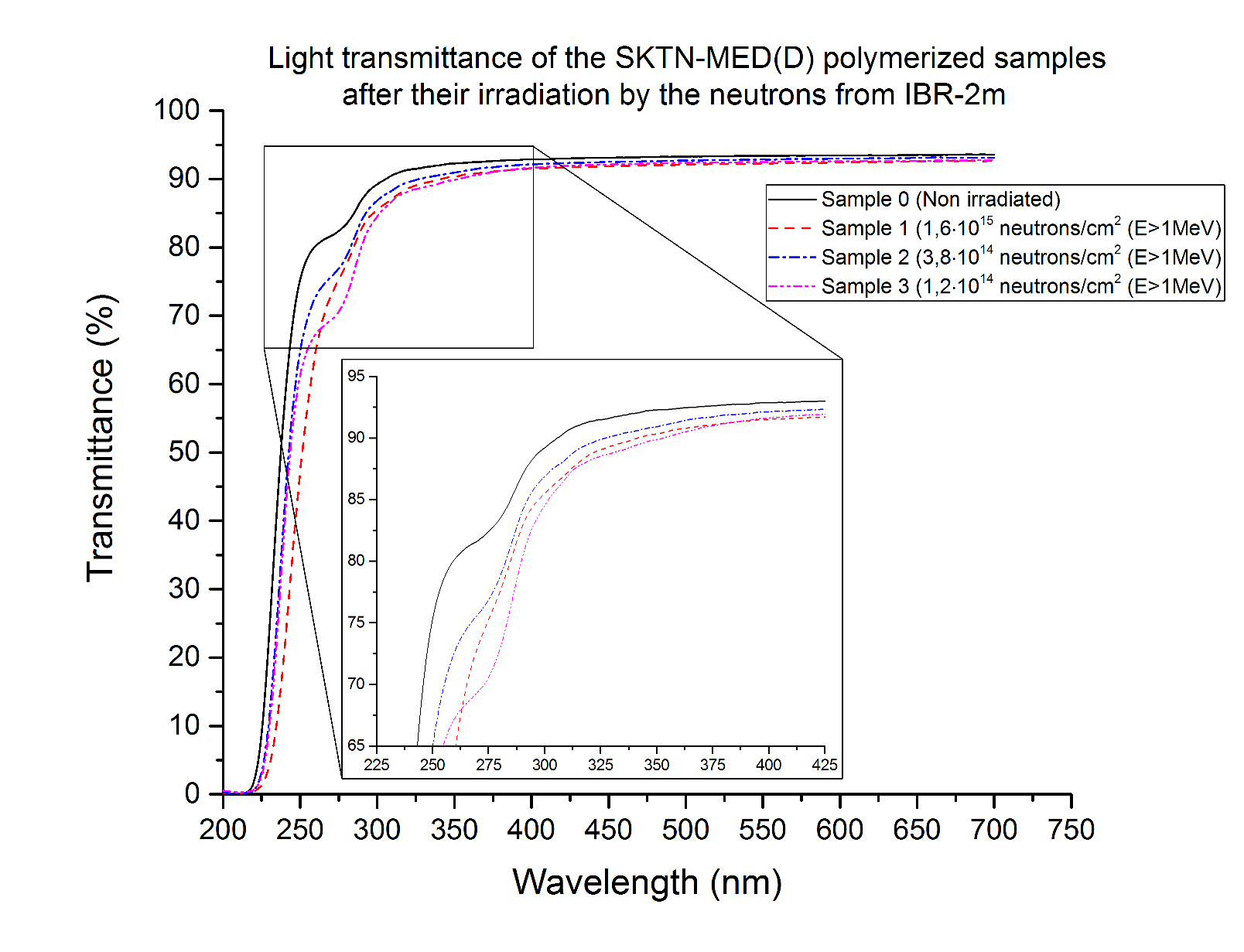}
%
		\caption{\small Transmittance of the sample (sheet) made of the SKTN-MED(D) glue before and after the irradiation by the neutron beam with the fluences of $16\times10^{14}$, $3.8\times10^{14}$, and $1.2\times10^{14}$ neutrons per cm$^{2}$.}
		\label{fig: transpSKTNhh}
	\end{center}
\end{figure}

It was of interest to study transmittance of the SKTN-MED(D) glue (resin+hardener). We made 4-mm-thick sheets using the SKTN-MED(D) glue. Only 2-6\% of hardener are needed for polymerization, which proceeds pretty fast (for about 30~min). Transmittances of these sheets were measured before and after the neutron irradiation ($16\times10^{14}$, $3.8\times10^{14}$, and $1.2\times10^{14}$ neutrons per cm$^{2}$; see Figure~\ref{fig: transpSKTNhh}). One can see that the transmittances are at a level of 90\% as high as the resin base transmittance (see Figure~\ref{fig: transpSKTN}). However, the actual transmission spectrum of the glue starts from about 225~nm. Note that this shift of the transmission spectrum is due to the fact that the Plexiglas windows do not transmit ultraviolet light with wavelengths longer than 300~nm.

\begin{figure}[!htbp]
	\centering
	\subfloat[BC-600 base]{\includegraphics[height=0.15\textheight, keepaspectratio]
		{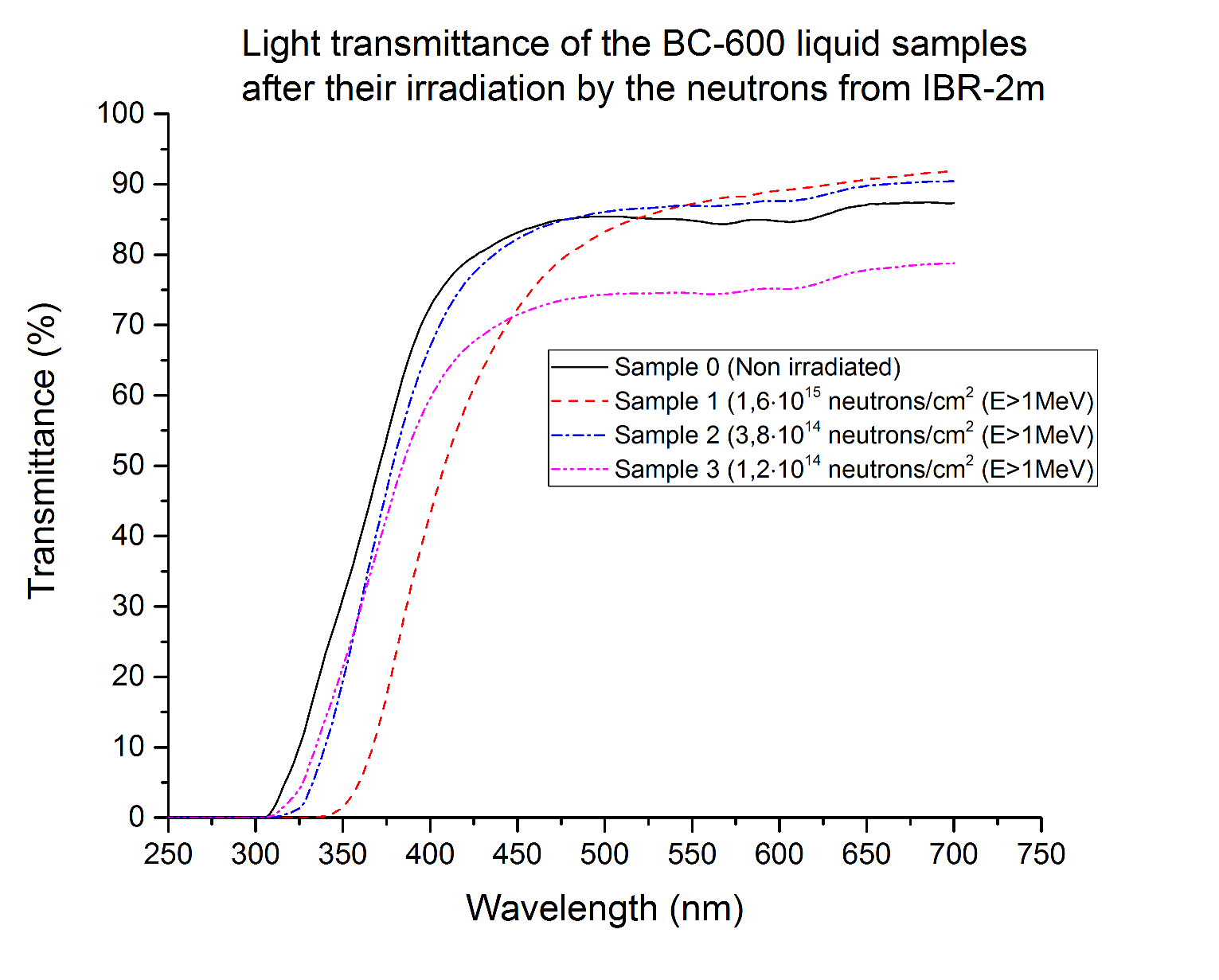} \label{fig: transpBC600base} }
	\hfill
	\subfloat[ВС-600 glue]{\includegraphics[height=0.15\textheight, keepaspectratio]
		{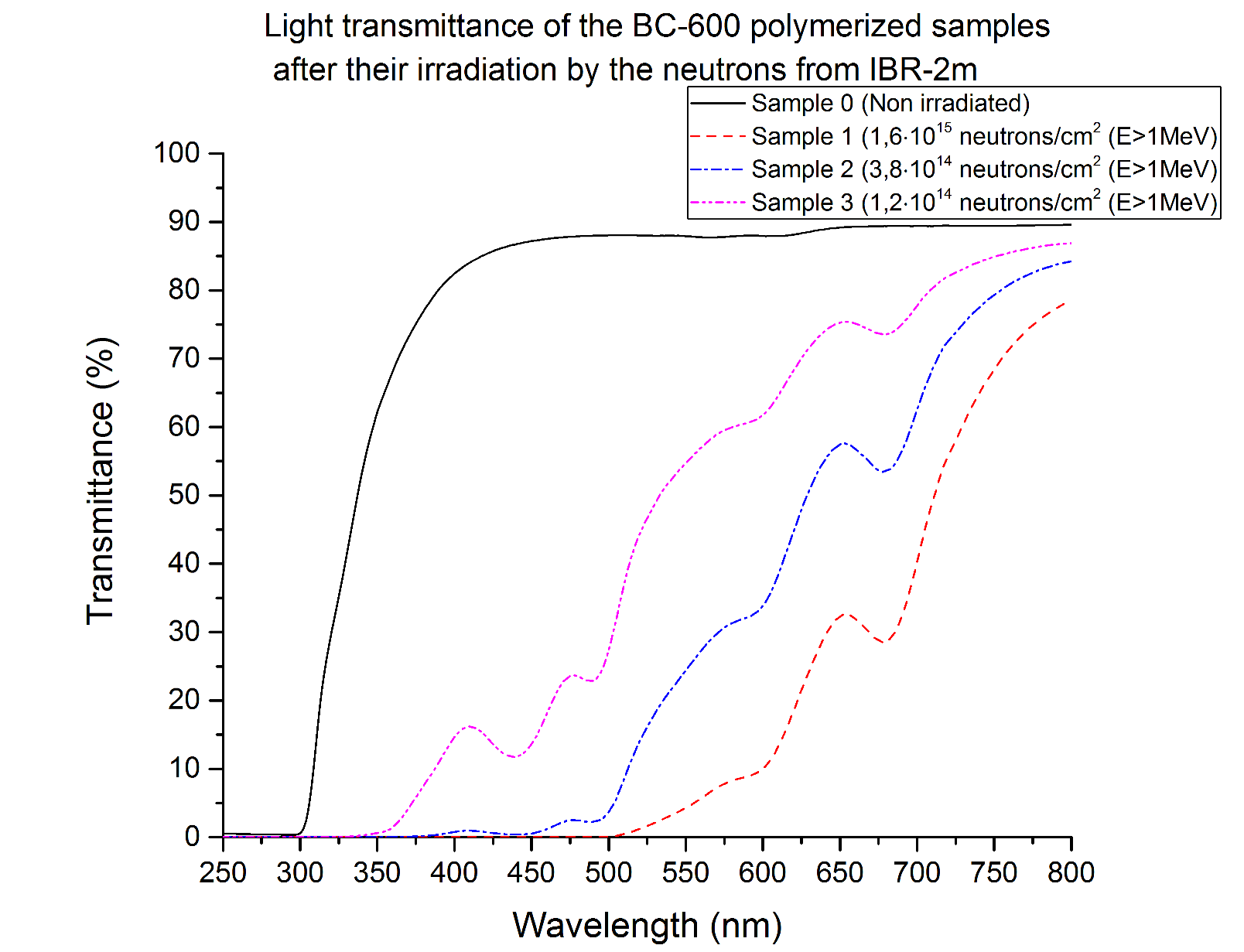} \label{fig: transpBC600glue} }
	\hfill
	\subfloat[ВС-600 glue]{\includegraphics[height=0.15\textheight, keepaspectratio]
		{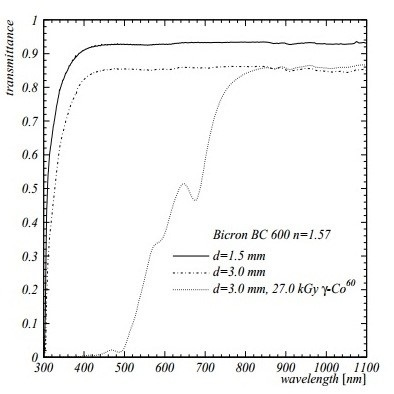} \label{fig: radinvest} }
	\caption{\small Transparency of the BC-600 base (a) and the BC-600 glue (b) before and after the irradiation by the neutron beam with the fluences of $16\times10^{14}$, $3.8\times10^{14}$ and $1.2\times10^{14}$ neutrons per cm$^{2}$. \\(c) Transmittance of the ВС-600 glue samples with various thickness $d$ before and after the irradiation \cite{radcms}.}
	\label{fig: transpBC600}
\end{figure}

It is also of interest to study radiation hardness of the widely used BC-600 optical epoxy glue and compare the results with those for SKTN-MED. We measured transmission of polymerized BC-600 glue and unpolymerized BC-600 base samples. The results for the BC-600 glue and its base before and after the irradiation by the  neutron beam with the fluences of $16\times10^{14}$, $3.8\times10^{14}$, and $1.2\times10^{14}$ neutrons per cm$^{2}$ are shown in Figures \ref{fig: transpBC600base} and \ref{fig: transpBC600glue} respectively. 

As expected, the BC-600 glue (and its base as well) was sensitive to the impact of the applied radiation dose rate (we can compare these results with \cite{radcms}). As described in this article, losses of transmission were studied in thin samples of glues with the thickness of 1.5 and 3.0 mm. The ВС-600 samples were irradiated with $\gamma$ particles emitted by $^{60}Co$ with the dose rate of 27 kGy. The results of the BC-600 radiation hardness study published in the article \cite{radcms} are presented in Figure \ref{fig: radinvest}. One can see good agreement between our results and the results obtained in \cite{radcms}.

It is worth noting that our study shows that polymerized BC-600 glue is more sensitive to radiation than its base without a hardener.

\begin{figure}[!htbp]
	\centering
	\subfloat[SKTN-MED(D)]{\includegraphics[height=0.1\textheight, keepaspectratio]
		{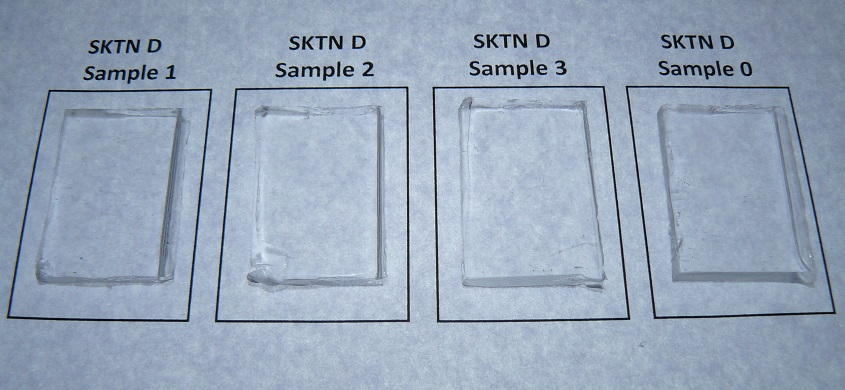} \label{fig: transpVisualSKTN} }
	\hfill
	\subfloat[ВС-600]{\includegraphics[height=0.1\textheight, keepaspectratio]
		{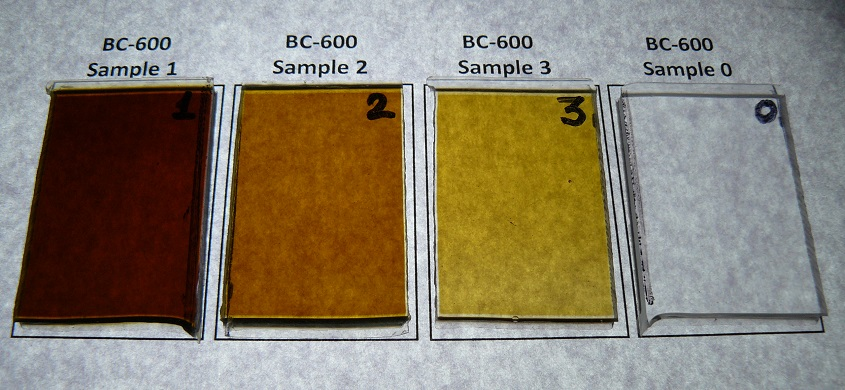} \label{fig: transpVisualBC600} }
	\caption{\small Photos of the SKTN-MED(D) and ВС-600 sheets irradiated by the neutron beam with the fluences of $16\times10^{14}$ (samples~1), $3.8\times10^{14}$, (samples~2), and $1.2\times10^{14}$ (samples~3) neutrons per cm$^{2}$. Samples 0 were not irradiated.}
	\label{fig: transpVisual}
\end{figure}

The photos of the SKTN-MED(D) and ВС-600 sheets irradiated with different dose rates are shown in Figure \ref{fig: transpVisual}.

\subsection{Changes in the light yield of the strips with and without a filler after irradiation on the IBR-2}\label{irbstrip}
It is of interest to study how the radiation affects optical properties of the strips with a filler and WLS fiber. For this purpose, the 15-cm-long strip samples with and without a filler were irradiated on the IBR-2 by fast neutron ($E>1 \: MeV$) fluxes with the fluences of $16\times10^{14}$, $3.8\times10^{14}$, and $1.2\times10^{14}$ neutrons per cm$^{2}$. The light yield of "dry" strips and strips filled with various resins was measured using a picoammeter and radioactive sources before and after the irradiation. The radioactive source was placed in the middle of the sample. The results are given in Table \ref{table: radcur}.

\begin{table}[!htbp]
	\footnotesize
	\begin{center}
		\begin{tabular}{|c|c|c|c|c|c|c|}
			\hline
			Sample& Sample& Fluence,& Filler& PMT anode& PMT anode& Decrease of\\
			number& name	& $10^{14}$ neutrons& &current& current & the anode\\
			& 			& per cm$^2$&			& before	& after	&	current after\\
			& 			& ($E>1MeV$)&			& irradiation	&	irradiation	&	irradiation					\\
			& 			& 			&			&$I_{b}$, nA	&	$I_{a}$, nA&	in \%					\\
			\hline
			1.1	& Dry 1	& 16 & None &	102(7) & 14(3) & 86(4) \\
			1.2	& SKTN(E) 1&	&	SKTN(E)& 126(9)& 18(3)&	86(3) \\
			1.3	& SKTN(D) 1&	&	SKTN(D)& 147(10)&	22(3)& 85(3) \\
			1.4	&BC-600 1	&	& BC-600&	141(10)&	20(3)&	86(3) \\
			\hline
			2.1	&Dry 2	&3.8	&None&	65(5)&	13(2)&	80(4) \\
			2.2	&SKTN(E) 2&		&SKTN(E)& 	144(10)&	81(6)&	44(6)\\
			2.3	&SKTN(D) 2&		&SKTN(D)	&166(11)	&55(4)	&67(4)\\
			2.4	&BC-600 2&		&BC-600	&149(10)	&80(6)	&46(6)\\
			\hline
			3.1	&Dry 3	&1.2	&None	&89(7)	&64(5)	&28(8)\\
			3.2	&SKTN(E) 3&		&SKTN(E)& 	148(10)&	105(8)	&29(8)\\
			3.3	&SKTN(D) 3&		&SKTN(D)&	147(10)&	103(8)	&30(8)\\
			3.4	&BC-600 3&		&BC-600&	151(10)&	86(7)	&43(6)\\
			\hline
		\end{tabular}
	\end{center}
	\caption{\small The light yield of "dry" strips and strips filled with various resins under the effect of the radioactive source before and after the irradiation of the strips by a neutron beam.}
	\label{table: radcur}
\end{table}

Absolute values of anode currents for strips with different fillers were quite similar before the irradiation, while after the irradiation they decreased in accordance with increasing neutron fluence. Since optical transmittance of SKTN-MED (D and E) practically did not change at those fluxes, light yield deterioration is mainly caused by destruction of the strip and the WLS fiber.

\begin{figure}[!htbp]
	\centering
	\subfloat[]{\includegraphics[height=0.15\textheight, keepaspectratio]
		{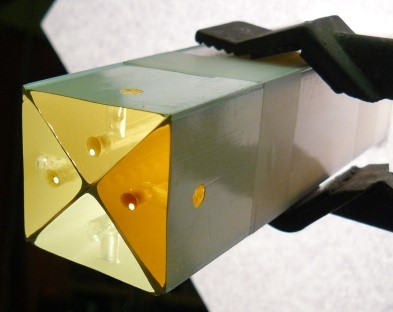}}
	\hfill
	\subfloat[]{\includegraphics[height=0.15\textheight, keepaspectratio]
		{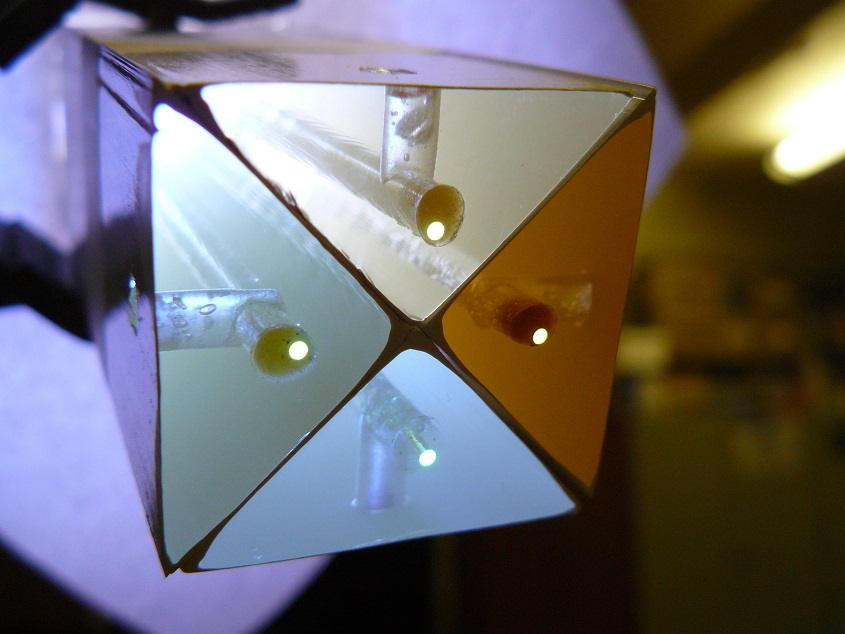}}
	\caption{\small Unexposed strip is located in the lowest position, followed clockwise by three other strips irradiated by the neutron beam with the fluences of $1.2\times10^{14}$, $3.8\times10^{14}$, and $16\times10^{14}$ neutrons per $cm^{2}$ respectively. The left photo was taken under the illumination by a regular incandescent lamp and the right photo was taken under blue light.}
	\label{fig: radstrip}
\end{figure}

Photos of four strips are shown in Figure \ref{fig: radstrip}. The unexposed strip is in the lowest position, followed clockwise by three other strips irradiated by the neutron beam with the fluences of $1.2\times10^{14}$, $3.8\times10^{14}$, and $16\times10^{14}$ neutrons per $cm^{2}$ respectively. One can see noticeable changes in the transmission of the irradiated strips in comparison to the unexposed strip as the absorbed dose rate increases. Degradation of plastic scintillator transmittance in the blue light region due to the absorbed radiation dose rate is visible under exposure to the blue light of the regular incandescent lamp. It is also clearly seen that the Hardman RED 04001 optically transparent 5-min glue based on epoxy resin completely lost transparency in this region of the spectrum.

\section{Conclusions}\label{conclusions}
\begin{enumerate}
	
	\item Applicability is shown of the developed technology for injection of a highly viscous optical filler, particularly SKTN-MED and BC-600, into the 5-m-long polystyrene strip with a co-extruded hole and with a WLS fiber embedded in it.
	
	\item Light yield study for the 5-m-long strip filled with SKTN-MED(E) carried out with cosmic muons and radioactive sources shows an increase in the light yield by a factor of up to 1.5 in comparison to the light yield of the "dry" strip when the far-from-PMT ends of the strip and the WLS fiber are blackened.

	\item SKTN-MED (D and E) and BC-600 as well have a good transparency (above 90\%) in a wide range of the light spectrum ($\lambda>400nm$), but only SKTN-MED holds it under the exposure to a neutron flux (including neutrons with $E>1\:MeV$) at fluences up to $3.8\times10^{14}$ neutrons per cm$^{2}$. This result allows using SKTN-MED(D) as a filler for the 6.6-m-long strips in the mu2e experiment, where radiation dose rates of $10^{11}$ neutrons per cm$^2$ \cite{mu2e, mu2e_oks} are expected for these long detectors during the operation time.

	\item  Damage of the strip and the WLS fiber is the main cause of light yield losses in strips with the SKTN-MED optical filler after their irradiation by neutrons with fluences up to $16\times10^{14}$ neutrons per $cm^{2}$.
	
	\item MC simulation using Geant~4 shows an increase in the light yield of the 5-m-long strip with a filler and generally agrees with the experimental data.

\end{enumerate}

\section{Acknowledgments}\label{thanks}
We thank our colleagues who provided insight and expertise that greatly assisted the research.

We also thank the graduate students A.~Boykov and I.~Zimin (FNIS, Dubna State University, Russia) for their helpful support.

We are grateful to A.~Zagartdinov (Surel Ltd.) for the technical support and consultation concerning the use of the low-molecular synthetic resin SKTN-MED.

We are grateful to S.V.~Morzhukhina (Ph.D in Chemistry, Associate Professor in Chemistry, Head of the Department of Chemistry, New Technologies and Materials, Dubna State University, Russia) and to I.V.~Mushina (Senior lecturer, Department of Chemistry, New Technologies and Materials, Dubna State University, Russia) who provided measurements of the refractive indices for different fillers.

We would also like to show our gratitude to M.~Potapov (DLNP, JINR) for language editing and proofreading.

\noindent\rule{\hsize}{0.4pt}
\newpage
\singlespacing

\addcontentsline{toc}{section}{References}\label{library}
\small


\end{document}